\documentstyle[aps]{revtex}

\def\be{\begin{equation}}
\def\ee{\end{equation}}
\def\ba{\begin{eqnarray}}
\def\ea{\end{eqnarray}}
\def\half{{1 \over 2}}
\def\tE{{\tilde E}}
\def\A{{\bf A}}
\def\E{{\bf E}}
\def\B{{\bf B}}
\def\x{{\bf x}}
\def\k{{\bf k}}
\def\vel{{\bf v}}
\def\a{{\bf a}}
\begin{document}
\input epsf
\draft
\twocolumn[\hsize\textwidth\columnwidth\hsize\csname
@twocolumnfalse\endcsname
\title{Induced vortex tunneling in a superconducting wire}
\author{S. Khlebnikov}
\address{Department of Physics, Purdue University, West Lafayette, 
IN 47907, USA}
\maketitle
\begin{abstract}
We consider induced topological transitions in a wire made from 
cylindrical superconducting film. During a transition, a pulse of electric 
current causes transport of a virtual vortex-antivortex pair around
the cylinder. We consider both the instanton approach, in which the transition
is viewed as motion of vortices in the Euclidean time, and the real-time
dual formulation, in which vortices are described by a fundamental quantum 
field. The instanton approach is convenient to discuss effects of the environment, 
while in the dual formulation we show that there exists a potentially useful 
adiabatic regime, in which the probability to create a real 
vortex pair is exponentially suppressed, but the total transport of the vortex
number can be of order one.
\end{abstract}
\pacs{PACS numbers: 85.25.Hv, 03.67.Lx}
\vskip2pc]
\section{Introduction}
Tunneling of magnetic flux across weak superconducting links of various 
nature is of interest for both fundamental physics and applications.
In particular, devices using Josephson junctions have been discussed
recently in connection with proposals for quantum computing (qubits).
(For a review
of various types of superconducting qubits, see ref. \cite{review}.)
Such devices (SQUIDs) can be, for 
many purposes, reduced to just a few degrees of freedom, so it becomes
possible to numerically simulate their quantum behavior. In this way, 
one can discuss not only the amplitude transfer between the basis states, 
but also the residual excitation due to population of the higher levels 
\cite{tipping,deex}.

As the strength of the weak link is increased, by going from
a Josephson contact to a narrow wire, and then to a wider wire, 
superconductivity becomes more robust, and the tunneling rate drops.
In ultrathin wires, with widths $L_x \ll \xi$, where $\xi$ is 
the Ginzburg-Landau coherence length, the relevant tunneling events are
quantum phase slips \cite{thin1,thin2}. 
Crossover to the wide-wire case occurs at widths $L_x \sim \xi$,
and for wider wires transport of flux happens via formation of vortices. 

Robustness of superconductivity in wires with $L_x \gg \xi$ can be useful
in applications that require long coherence time, such as qubits.
Indeed, we expect that in a
suitable implementation such a wire will not be subject to any shunting
effects except for thermal quasiparticles.
Possible implementations include a wire defined lithographically on
a substrate or a multiwall carbon nanotube coated with a superconducting
film. In either case, the superconducting circuit has to be closed to allow a
persistent current, which presumably can be done using ordinary bulk
superconductors.

On the other hand, when a wire with $L_x \gg \xi$
is used in place of a weak link,
inducing a transition between the basis states of the qubit becomes
a nontrivial matter, since flux does not travel easily across wider wires.
We propose to use for this purpose a pulse of electric current 
along the wire, which can be produced by coupling the qubit, either directly
or inductively, to some external circuit. The pulse
will lower the potential barrier separating the basis states and thus 
encourage the motion of flux across the wire. We call this process
{\em induced} vortex tunneling.\footnote{As we explain 
in more detail below, the pulse of current is 
supposed to enhance tunneling of flux but {\em not} induce real-time, 
over-barrier transitions. Thus, our problem is quite distinct from that
of a resistive state in thin wires \cite{Galaiko&Kopnin} or films 
\cite{Weber&Kramer}. In particular, real-time classical 
simulations based on the time-dependent Ginzburg-Landau equation 
are clearly inapplicable in our case.} 

Because this process involves many degrees of freedom, the theory of it
has to be constructed from
an entirely different standpoint than the few-degree-of-freedom modeling 
common in the theory of SQUIDs. In this paper we develop and compare two
approaches. One approach uses instantons, i.e., Euclidean solutions, which
in our case correspond to motion of vortices and antivortices across the 
wire in Euclidean (imaginary) time. This approach makes it possible to 
discuss the effects of the environment (specifically, fermions at the 
vortex cores) on tunneling, but for calculations of the residual excitation
left in the system by the pulse, is less convenient than the other approach, 
based on duality. In the dual formulation, vortices are described by a 
fundamental quantum field, and the tunneling rate appears as a 
finite-size effect due to mode quantization in a confined geometry.

For definiteness, we consider cylindrical geometry, in which the wire 
is formed by a thin
superconducting film on an insulating cylindrical surface. The ``width''
$L_x$ of the wire is the circumference of the cylinder. Although this may 
not be the simplest geometry to manufacture, it has the advantage of simple 
(periodic) boundary conditions in the $x$ direction. 
Results for other geometries, e.g., thin strips should be substantially 
similar.

\begin{figure}
\leavevmode\epsfysize=2.0in \epsfbox{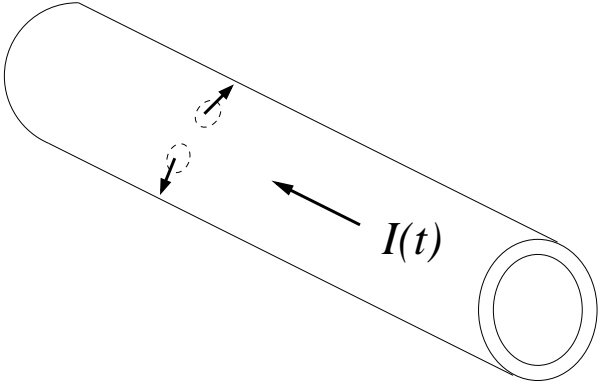}
\vspace*{0.1in}
\caption{A segment of wire made from cylindrical superconducting film. 
A pulse of current $I(t)$ 
induces transport of a virtual vortex-antivortex pair around the cylinder.
}
\label{fig:cyl}
\end{figure}

Short Abrikosov flux lines that can propagate in the film will be
referred to as vortices and antivortices, depending on which direction the
flux points.
Motion of the flux across a cylindrical wire can be viewed as creating
a vortex-antivortex pair on one side of the cylinder, 
transporting them along the circumference, and annihilating them
on the other side, see Fig. \ref{fig:cyl}. 
This process has been discussed in various contexts in 
the literature \cite{Wen&Niu,Kitaev,thermal}. Here, we present a detailed
calculation of the rate, including the effect of core fermions, and 
a discussion of the residual excitation.

Vortex instantons have been used to discuss the phenomenon of quantum creep 
\cite{creep}. That discussion employed the ordinary non-relativistic vortex
action. In our case, since a tunneling event requires formation of 
a vortex-antivortex pair, we are constrained to use a ``relativistic'' 
action, with the speed of light replaced by some other limiting speed $c_1$.
We will see that in this ``relativistic'' case, instantons are characterized 
by large (Euclidean) velocities and correspondingly short timescales, one of
the consequences being that the core contribution to the vortex mass is
greatly reduced.

We stress that when we speak about production of a vortex pair,
transporting vortices around the cylinder, etc. we mean a {\em virtual} process,
i.e., tunneling. Production of real vortices is detrimental to our
goal, since these vortices would be easily ``detected'' by the environment
(e.g., by electrons at the vortex cores), and that would result in rapid
decoherence of quantum superpositions that we intend to form. To prevent
decoherence, induced transitions should leave as little imprint on the
environment as possible. The main result of the present work is that
it is possible, at least theoretically, to use a slowly (adiabatically)
changing current that has an exponentially small probability to create
a real vortex but still leads to a sizeable transport of the vortex number
around the cylinder.

At the first-quantized
level, the dual description we use coincides with that employed in the theory
of vortex transport in arrays of Josephson junctions \cite{arrays}. 
However, to consider effects of  
zero-point vortex-antivortex pairs fluctuating ``in and out of existence'',
we need to apply duality at the second-quantized level. 
In a second-quantized description, vortices are represented by a fundamental
quantum field. Such a description was used in refs. \cite{Ao,IJ} 
to study nucleation of vortex-antivortex pairs in
the presence of a static supercurrent. 
The main differences between that problem and ours
are that in our case the process is, first, time-dependent
and, second, avoids as much as possible production of real vortex pairs.

The paper is organized as follows. Sect. II describes the duality map. 
Effects of core fermions and of the vortex-antivortex potential
on vortex tunneling are discussed,
using the instanton picture, in Sect. \ref{sect:env}. This discussion gives 
us confidence that the linear equation (\ref{eqm}) obtained in Sect. 2 indeed
captures the essential physics of the process.
Sect. IV contains definitions of the average vortex current, which is used 
to measure the tunneling amplitude, and of the residual 
excitation left in the system after the pulse (the number of real 
vortices  produced). In Sect. V we consider the adiabatic limit (a
slowly changing current) and show that it is
possible to have a sizeable amplitude transfer with exponentially small 
residual excitation. Some numerical estimates are given
in Sect. \ref{sect:num}. We find that for a film of thickness $d=10$ nm, 
a suitable width of the wire is of order 1 $\mu$m. Sect. VII is a conclusion.

\section{Duality map}
The original quantum description of a superconductor contains the usual
vector potential $A_{\mu}$ as well as fields corresponding to other degrees
of freedom: fluctuations of the order parameter, normal electrons at vortex
cores, etc. We introduce field $a_{\lambda}$, dual to the electromagnetic 
field $A_{\mu}$, 
by rewriting the corresponding factor in the functional integral
as a Gaussian integral:
\ba
\lefteqn{\exp\left( - \frac{id}{16\pi\hbar} \int F_{\mu\nu}^2 d^2x dt\right) 
= {\rm const.} \times } 
\nonumber \\
 & & \int {\cal D} a_{\mu} \exp \frac{i}{\hbar} \int
\left( \frac{gc}{4\pi} \epsilon^{\mu\nu\lambda} F_{\mu\nu} a_{\lambda}
+ {g^2 c^2 \over 2\pi d} a_{\lambda}^2 \right) d^2x dt \; .
\label{p-by-p}
\ea
Here $d$ is the film's thickness, 
\be
g = \frac{2e}{\hbar c} \; ,
\label{g}
\ee
$e$ is the magnitude of electron charge ($e>0$), 
and $c$ is the speed of light.
We have already incorporated the condition
that the film is thin, so that parallel fields penetrate inside undiminished.
Greek indices run over values 0,1,2; Latin indices over 1,2.
Sums over repeated indices are implied.

The next step is to separate the original field $A$ into $A^{\rm (vort)}$, which
consists of narrow flux tubes attached to vortices, and the remaining long-range 
part $A^{\rm (long)}$. In thin films, the screening
length for perpendicular magnetic fields is very large \cite{Pearl}, so a
narrow flux tube does not account completely for the magnetic field of a vortex:
part of the field will appear in $A^{\rm (long)}$ and will be responsible 
for long-range interactions between the vortices. However, the electric field
of a moving vortex is confined to the vortex core \cite{Suhl,Sonin&al,thermal},
and that makes introduction of the flux tubes useful. 
For the field strength we have
\be
F_{\mu\nu} = F_{\mu\nu}^{\rm (long)} + F_{\mu\nu}^{\rm (vort)} \; .
\label{split}
\ee
We define a (conserved) current of vortices as
\be
J^{\lambda} 
\equiv - \frac{gc}{4\pi} \epsilon^{\mu\nu\lambda} F_{\mu\nu}^{\rm (vort)} \; .
\label{Jdef}
\ee
Thus a state with $J_0 = c \delta(\x)$ carries 
precisely a unit flux quantum $\Phi_0 = 2\pi / g$ of $F^{\rm (vort)}$.

To obtain a dual description of vortices, we introduce a new complex 
field $\chi$ with a unit charge with respect to the field $a$.
The action of the vortex field is taken in the form
\ba
\lefteqn{ S_{\chi} = \int d^2 x dt \left[ 
\left| \partial_t \chi + (ic/\hbar) a_0 \chi\right|^2 \right.}
\nonumber \\
& & \left. - c_1^2 \left| \partial_i \chi + (i/\hbar) a_i \chi\right|^2
 -  M^2(t) |\chi|^2 \right] \; ,
\label{Schi}
\ea
where $M(t)$ is half of the pair-production frequency. During a pulse of current,
this frequency is time-dependent.
Eq. (\ref{Schi}) is a ``relativistic'' action, except that the role of the speed
of light is played by some other speed $c_1$, cf. refs. \cite{Ao,IJ}. 
We will discuss the value of $c_1$ in Sect. \ref{sect:num}.

The ``relativistic'' form of (\ref{Schi}) may seem ad hoc, but in fact
the form of the vortex action is highly constrained by symmetries. For
example, the presence of a conserved vortex current dictates 
that $\chi$ should be a complex field, while the possibility of pair 
production dictates that the kinetic term should be quadratic in the
time derivative.
The absence of terms with higher spatial derivatives is 
a good approximation as long as wavenumbers do not exceed a certain 
ultraviolet cutoff, which we expect to be of order of the lattice spacing.
Finally, self-interaction of the
vortex field (higher-order terms in $\chi$) can be neglected as long as 
vortices are sufficiently rare.

The simple action (\ref{Schi}) neglects other degrees of freedom, 
in particular, fermions at the vortex core \cite{CdGM}. 
These are nearly gapless and, as well known,
cause large friction for slow vortex motions \cite{friction}. For a tunneling
process, involving virtual vortices only, the expected effect of such 
a dissipative
environment is a reduction in the tunneling rate \cite{CL}. 
We analyze this effect in the
next section and find that for rapid, large-frequency Euclidean motions
corresponding to tunneling, the role of dissipative processes is greatly reduced,
so much so that they do not affect the main exponential factor in the tunneling
rate, but can only affect the preexponent.
Because our main conclusions are based on the value of the exponential factor,
we take 
\be
S_{\rm dual} = \int d^2x dt \left( \frac{gc}{4\pi} \epsilon^{\mu\nu\lambda} 
F_{\mu\nu}^{\rm (long)} a_{\lambda} 
+ {g^2 c^2 \over 2\pi d} a_{\lambda}^2 \right) + S_{\chi} 
\label{Sdual}
\ee
as the full action of our dual description.

The vortex current in this description is given by the usual formula
$J^{\lambda}=-\delta S_{\chi}/\delta a_{\lambda}$. In particular, the
spatial components are
\be
J^i =  - \frac{i}{\hbar} c_1^2 \left( \chi^{\dagger} \partial_i \chi - 
\partial_i \chi^{\dagger} \chi + \frac{2i}{\hbar} \chi^{\dagger} \chi a_i
\right)  \; .
\label{Ji}
\ee
From (\ref{Sdual}), we obtain an equation for $a$ in the form
\be
{g^2 c^2 \over \pi d} a^{\lambda} = 
- \frac{g c}{4\pi} \epsilon^{\mu\nu\lambda} F_{\mu\nu}^{\rm (long)} 
+ J^{\lambda} \; .
\label{eqa}
\ee
Combined with eqs. (\ref{split}) and (\ref{Jdef}) this is seen to be
completely equivalent to the saddle-point condition for the integral
(\ref{p-by-p}), i.e., to the original definition of $a$.

Equation of motion for $\chi$ is obtained from (\ref{Schi}) and reads
\be
[\partial_t + (ic /\hbar) a_0]^2\chi
- c_1^2 [\partial_i + (i/\hbar) a_i]^2 \chi + M^2 \chi = 0 \; .
\label{eqm}
\ee
When we use (\ref{eqa}) in this equation, we obtain an interaction of
vortices with the long-range field $F_{\mu\nu}^{\rm (long)}$. (The interaction
with the short-ranged $J^{\lambda}$ is negligible if vortices are rare.)
As we have mentioned
above, the field $F_{\mu\nu}^{\rm (long)}$ includes the long-range tails 
\cite{Pearl} of vortex magnetic fields and thus gives rise to a logarithmic
interaction between vortices. Accordingly, the frequency $M$ in (\ref{eqm}) 
should be understood as being due to the vortex core only. However,
for instanton processes, involving a single vortex-antivortex pair, we can,
instead of
keeping track of the vortex contribution to $a^{\lambda}$, model the
vortex-antivortex potential by making $M$ position-dependent.
Moreover, as we will see in the next section, 
for calculation of the tunneling exponent,
such $x$-dependent frequency can be replaced simply by its average value 
along the vortex path. Within this model,
and with the short-ranged $J^{\lambda}$ dropped from (\ref{eqa}), components 
of $a$ do not contain any fields due to vortices and reduce to external 
$\E$ and $\B$ fields only:
\be
(a_0,a_1, a_2) = \frac{d}{2gc} (B_z, -E_y, E_x) \; .
\label{acomp}
\ee

Eq. (\ref{eqm}) is a second-quantized description of vortices, which
allows us to study effects due to virtual vortex-antivortex pairs.
Similar descriptions were used previously in refs. \cite{Ao,IJ} to study the 
instability of a static supercurrent via production of real 
vortex-antivortex pairs. In the limit when vortex-antivortex fluctuations
are neglected ({\em not} a suitable limit here), eq. (\ref{eqm}) 
gives rise to a first-quantized description, 
in which a vortex of mass $m = \hbar M/c_1^2$ 
moves under action of a Lorentz force built from dual electric
and magnetic field. These fields, 
${\bf e} = -\nabla a_0 - (1/c) \partial_t \a$ and
${\bf b} = \nabla\times \a$ can be expressed through the electric
current and charge density, using (\ref{acomp}) and Maxwell's equations.
The resulting expressions coincide with those appearing in the 
first-quantized theory of ref. \cite{arrays}.

For a static supercurrent $I$ in the $y$ direction, 
we would have $a_1(t) \propto I t$. In contrast, the adiabatic case considered below
corresponds to a slowly changing pulsed $a_1$: $a_1(t\to\pm\infty) = 0$.

An additional constant ``magnetic'' field ${\bf b}$ can be used to describe a Magnus 
force on the vortex. However, as we will find in the next section, 
for tunneling processes the Magnus force is canceled, to a very high accuracy, 
by the spectral flow force \cite{KK,Volovik,Stone} due to core fermions.
Accordingly, we set the total effective Magnus force to zero: ${\bf b} = 0$.

Eq. (\ref{eqm}) simplifies further if we restrict our attention to the
cylindrical geometry of Fig. \ref{fig:cyl}.
A pulse of electric current will be applied lengthwise, in the $y$ direction.
The resulting field $E_y$ will induce a virtual pair of vortex
and antivortex to travel along the circumference of
the cylinder (i.e., in the $x$ direction). So, we set
$E_x= B_z=0$, and $E_y = E(t)$. For the
present configuration, $E$ is independent of $x$ and $y$.
Thus, the equation of motion takes the form
\be
\ddot{\chi} - c_1^2 \partial_y^2 \chi 
- c_1^2 \left[ \partial_x - i \frac{d}{4e} E(t) \right]^2 \chi
+ M^2(t) \chi = 0 \; .
\label{eqm2}
\ee
Although we will not need any such relation in what follows, we note that
the two time-dependent parameters 
in (\ref{eqm2}), $E$ and $M$, can be expressed
through a single function of time---the electric
current (see ref. \cite{num} for details).

\section{Instantons and the effects of the environment} \label{sect:env}
It is well known that motion of vortices in superconductors is subject
to strong renormalization effects due to the ``environment'' of normal
electrons at the vortex cores. For low-frequency motions, in very clean
superconductors, the main effect is a drastic reduction of
the limiting speed $c_1$ in eq. (\ref{eqm}).
Indeed, in such cases it is natural to expect that $c_1$ will be equal
to the critical velocity of the superconductor $v_{\rm cr} = \Delta/v_F$,
where $\Delta$ is the value of the gap outside the core. At larger
speeds, the vortex becomes a ``tachyon'' and causes an instability
via production of quasiparticles from the core. Equivalently, the small
$c_1= v_{\rm cr}$ can be interpreted as a large value of the inertial
mass $m$ of the vortex, since from (\ref{eqm})\footnote{
In this section, we set $\hbar=1$.}
\be
m = M / c_1^2 \; .
\label{m}
\ee
This estimate is consistent with direct calculations \cite{mass}
of the inertial mass (provided the frequency $M$ is understood as the core 
contribution only).

Situation, however, is entirely different at large frequencies, or short
timescales, which
as we will see below is the case relevant to tunneling.
The characteristic
frequency associated with the fermionic response is determined by the 
``minigap'' $\omega_0 \sim \Delta^2 / \epsilon_F$, separating different
states at the vortex core \cite{CdGM}. 
When the timescale of the vortex motion is much 
shorter than $\omega_0^{-1}$, the response and the contribution of fermions
to the inertial mass will be diminished. Here, we calculate 
this effect for the vortex motion corresponding to our tunneling problem.
The motion takes place in the Euclidean (imaginary) time.
We will employ the kinetic equation \cite{Stone} for core states, 
which gives their
evolution in real time, and then analytically continue the results to the
Euclidean domain.

The low-lying (``anomalous'')
branch of fermion spectrum \cite{CdGM} at the vortex core
runs in energy approximately from $-\Delta$ to $\Delta$
(for a review and further references, see ref. \cite{Kopnin}). This branch is
characterized by two quantum 
numbers: $k_z$, the momentum in the $z$ direction (which in our case is 
perpendicular to the film), and the angular momentum $l$. For a vortex at rest,
the energy levels
are given approximately by $E_l(k_z) = - \omega_0(k_z) l$, with
$\omega_0(k_z) =  \Delta^2/2v_F k_\perp$, and $k^2_\perp = k_F^2 - k_z^2$.
In particular, $\omega_0(0) = \Delta^2 / 2v_F k_F$ \cite{Stone}.
Motion of the vortex relative to the superfluid results in an interaction
Hamiltonian and can cause transitions between different core levels.
The distribution
function $n$ in general depends on the quantum numbers $k_z$ and $l$
and their conjugate coordinates $z$ 
and $\phi$. If $n$ is independent of $z$, the kinetic equation \cite{Stone}
reads
\be
\frac{\partial n}{\partial t} 
- \omega_0(k_z) \frac{\partial n}{\partial \phi}
+ \k  \times \vel(t) \frac{\partial n}{\partial l} = 0 \; ,
\label{kin}
\ee
where $\k=(k_\perp \cos\phi,  k_\perp \sin\phi, k_z)$, and $\vel$ is the vortex
velocity.
Note that we write the kinetic equation in the collisionless (Vlasov) 
approximation. This is appropriate whenever the characteristic timescale of
the motion is much shorter than the collision time.

Solutions to eq. (\ref{kin}) can be obtained analytically
for any form of the vortex velocity. Here we consider the case of main 
relevance
to us, when the velocity is in the $x$ direction at all times. Then, the
zero-temperature solution is a step-function
\be
n(\phi, l; t) = \theta\left[
l + k_\perp \omega_0 \int dt' G_R(t-t') h(\phi; t') \right] \; , 
\label{sol}
\ee
where
\be 
h(\phi; t) = v(t) \cos\phi + \omega_0^{-1} \partial_t v(t) \sin\phi \; ,
\label{h}
\ee
and $G_R$ is the retarded Green function for $\partial_t^2 + \omega_0^2$:
$G_R(t<t') = 0$, $G_R(t>t') = \omega_0^{-1} \sin[\omega_0 (t-t')]$. 
(More generally, the step-function can be replaced by an arbitrary function 
in (\ref{sol}), so finite-temperature solutions can also be constructed.)
Eq. (\ref{sol}) describes excitation of the core system, originally in its
ground state, by the vortex motion.

If velocity $v(t)$ is peaked around $t=0$ at the timescale $t_1$ shorter 
than both $\omega_0^{-1}$ and the ``observation time'' $t$, we can replace
it in (\ref{h}) with a pulse of the form $v(t) = x \delta(t)$, 
where $x$ is the
displacement. This yields $n(\phi, l; t) = \theta(l)$ for $t< 0$, and
\be
n(\phi, l; t) = \theta\left[l + k_\perp x \sin(\omega_0 t + \phi) \right] 
\label{n-high}
\ee
for $t > 0$. Integrating over all modes, we obtain the total momentum
transferred to electrons upon a vortex displacement $x$:
\be
p_x(x) = {d\over 2} x \int_{-k_F}^{k_F} \frac{dk_z}{2\pi} k_\perp^2 
\sin\omega_0 t \; ,
\label{px}
\ee
while for $p_y$ the sine gets replaced by $\cos\omega_0 t$. The remaining
dependence on $t$ means that even after the displacement has been completed,
and the vortex has stopped, fermions at the core continue to oscillate
(``wobble'').

Tunneling in a wire can be visualized as motion in Euclidean time
of a vortex and an antivortex,
each by half-width of the wire, or alternatively as motion of the vortex 
alone by the full width $L_x$. 
For definiteness, we will discuss the latter
process. We refer to it as an instanton.
The instanton has two timescales associated with it: time $\tau_0$
it takes to create or destroy a vortex-antivortex pair, and time $\tau_1$
it takes the vortex to travel the distance $L_x$. 
Both $\tau_0$ and $\tau_1$ are variational parameters that adjust 
themselves to achieve the largest tunneling amplitude. 

A natural estimate for $\tau_0$ is $\tau_0 \sim M^{-1}$, and we will see that 
this estimate is unaffected by the presence of the environment.
We will also find that in the limit when instantons are dilute, 
i.e., when the vortex motion has a large Euclidean action, we have
$\tau_1 \gg M^{-1}$, but still $\omega_0 \tau_1 \ll 1$.
So, to understand the effect of the environment on $\tau_1$, we analytically
continue to $t = -i\tau$ and expand in small
$\omega_0\tau$. The change in the vortex momentum is opposite to (\ref{px}),
and the corresponding contribution to the Euclidean action of the vortex is
\be
i \int_0^{L_x} p_x dx = {d\over 16} L_x^2  k_F^3 \omega_0(0) \tau \; .
\label{cor1}
\ee
In our case, we need to replace $\tau$ with $\tau_1$,
and although this just 
invalidates the approximation under which (\ref{px}) was obtained, and the
precise form of the velocity pulse begins to matter, we can still obtain
an order-of-magnitude estimate for $\int p_x dx$. Let us first obtain this 
estimate for the case when the instanton gas just ceases being dilute,
and $\tau_1$ becomes of order $M^{-1}$. 
In the case of induced tunneling, parameters $M$ and $\Delta$ 
should be taken at time when the transition actually
occurs. However, because (\ref{cor1}) is now proportional to $\Delta^2/M$, and
$M$ scales as $\Delta^2$, we can as well use the unperturbed,
equilibrium values. Using $M\sim (k_F d) \epsilon_F$, and 
$\omega_0(0) = \Delta^2 /4\epsilon_F$, we estimate (\ref{cor1}) as
$(k_F L_x)^2 (\Delta /  \epsilon_F)^2 / 64$. This is small for typical values
$k_F L_x \sim 10^3$ and $\Delta / \epsilon_F \sim 10^{-3}$, but can become
large for larger $L_x$. In the latter case, or in the dilute regime when
$\tau_1$ is sufficiently large in comparison with $M^{-1}$,
the action (\ref{cor1}) begins to play a role in
determining the value of the variational parameter $\tau_1$.

To see how that happens, consider the Euclidean action for a single
vortex, without the environment correction:
\be
S_E = \int d\tau M (1 + v_E^2)^{1/2} \; ,
\label{SE}
\ee
where $v_E$ is the Euclidean velocity, and we use units with $c_1 = 1$.
Strictly speaking, the mass $M$ should be made $x$-dependent, to include
the vortex-antivortex potential. For now, however, we will consider it as
some constant average mass; we will justify this replacement later.
Then, during the vortex motion, $v_E$ is constant,
and $\tau_1 = L_x/ v_E$. The exponential factor in the tunneling amplitude
in this case is given by
\be
\int \frac{d\tau_1}{\tau_1} e^{-S_E} =
\int dv_E \exp \left[ - \frac{M L_x}{v_E} (1 + v_E^2)^{1/2} - 
\ln v_E \right] \; .
\ee
For large values of the product $M L_x$, the saddle-point value of $v_E$
is large: $v_E = (M L_x)^{1/2}$, while $\tau_1 = (L_x / M)^{1/2}$.
The saddle-point value of the exponent in this case is
\be
S_E = M L_x / c_1 + O(1) 
\label{saddle}
\ee
(we have restored $c_1$ in this formula), so the limit of large $ML_x$
is precisely the limit when the instanton gas is dilute. Now, including
the correction (\ref{cor1}), with $\tau \sim \tau_1$, we obtain a new
effective action, which in the limit of large $v_E$ has the form
\be
S_{\rm eff} \approx M L_x +\frac{M L_x}{2v_E^2} + 
C L_x^3  k_F^3 \omega_0(0) {d\over 16 v_E}   + \ln v_E \; ,
\label{Seff}
\ee
where $C$ is a numerical coefficient of order 1. 
Depending on the parameters, either the second or the third term in 
(\ref{Seff}) can be more important in determining the saddle point value
of $v_E$. In either case, however, the effective action on the saddle point
will have the same form as eq. (\ref{saddle}). We conclude that the longitudinal 
momentum transfer from the vortex to fermions at the core affects at most 
the tunneling preexponent, but not the main exponential factor.

The $\exp(-{\rm const.} L_x)$ form of the tunneling exponent had been
discussed in the literature for various systems supporting vortices
\cite{Wen&Niu,Kitaev}. The main point of this section is that it
holds also for the ``relativistic'' action (\ref{SE}), and in this 
case the value of the exponent is not affected by the longitudinal correction
(\ref{cor1}).

Turning to the transverse momentum, and following the steps that led to
eq. (\ref{cor1}), we obtain the following
contribution to the Euclidean action:
\be
i p_y(L_x) y 
= i \frac{k_F^3 d}{3\pi} L_x y \left[ 1 + O(\omega_0^2 \tau_1^2) \right] \; ,
\label{cor2}
\ee
where $y$ is a small displacement in the $y$ direction.
Unlike (\ref{cor1}), this contribution is purely imaginary: it gives rise
to phase differences between vortex paths lying at different values of the
lengthwise coordinate $y$. If significant phase differences existed, the
tunneling amplitude would be drastically reduced. However, the magnitude of
the leading term in (\ref{cor2}) is such that, for $k_F$ equal to the value
of the Fermi momentum outside the vortex core, it precisely cancels the action 
corresponding to the Magnus force. This is the well known cancellation of 
the Magnus force by the force due to spectral flow 
\cite{KK,Volovik,Stone,micro,FGLV}. The remaining correction in
(\ref{cor2}) is of order $\omega_0^2$ and is already quite small. Still,
an even more complete cancellation can be achieved, because $k_F$, which 
measures the electron density at the core, is also a variational parameter
and does not have to coincide with the value outside the core.
As eq. (\ref{cor2}) shows, a tiny self-adjustment of $k_F$ is sufficient
to render the effective Magnus force exactly zero.

Finally, we need to restore the effect of the vortex-antivortex potential.
As before, we consider the situation when, upon creation of a virtual
vortex-antivortex pair, the antivortex stays in place, while the vortex makes
a complete circle around the cylinder. Then, for the vortex action, we can still
use eq. (\ref{SE}) but $M$ should now be replaced by some (periodic) function
of $x$. Accordingly, $v_E$ is no longer a constant. We notice, however, that
the reason why (\ref{SE}) previously simplified into (\ref{saddle}) was not that
$v_E$ was constant, but that it was large: $v_E \gg 1$. This still applies now,
during most of the path. Taking this limit in eq. (\ref{SE}) with $M\to M(x)$,
we obtain
\be
S_E \approx \frac{1}{c_1} \int M(x) dx \; ,
\label{Mx}
\ee
which is equivalent to (\ref{saddle}) with $M$ replaced by its average value
along the path.

\section{Observables}
We have seen in the preceding section that the instanton picture of tunneling,
with instantons corresponding to vortices and antivortices moving along closed
paths in the Euclidean time, provides a very appealing and intuitive picture
of flux tunneling across a wire of width $L_x \gg \xi$.
This picture has allowed us to estimate the effects of the environment and to
obtain the main exponential factor in the tunneling amplitude. However, for
certain tasks, in particular for understanding the adiabatic limit, it is more 
convenient to use
the dual description (\ref{eqm}). In this section, we discuss what observables
can be calculated in that dual theory that will contain information about
tunneling.

We will be interested primarily in two observables. The first is the
average vortex current. The corresponding operator is given by
(\ref{Ji}); in our present case, only $J_x$ has a nontrivial average. To reduce
the number of factors of $\hbar$ in the subsequent formulas, it is
convenient to introduce instead of $\chi$ a new field
\be
X = {\chi \over \sqrt{\hbar}} \; .
\ee
Then,
\be
J_x =  - i c_1^2 \left[ X^{\dagger} \partial_x X - 
\partial_x X^{\dagger} X - \frac{id}{2e} E(t) X^{\dagger} X 
\right]  \; .
\label{Jx}
\ee

For our cylindrical configuration, the field $X$ satisfies periodic boundary
conditions in the $x$ direction. The boundary conditions in the $y$ direction
will not matter; we take them to be periodic as well.
The Fourier expansion of $X$ is 
\be
X(\x,t) = \sum_{\k} \left( \alpha_{\k} f_{\k}(t) 
+ \beta^{\dagger}_{-\k} f_{\k}^* \right) e^{i\k\x} \; ,
\label{four}
\ee
where $\k = (k_x, k_y)$,
$\alpha$ and $\beta$ are the usual annihilation operators,
and the mode functions satisfy
\be
\ddot{f_{\k}} + \omega_{\k}^2(t) f_{\k} = 0 
\label{f}
\ee
with a time-dependent frequency
\be
\omega_{\k}^2(t) = c_1^2 k_y^2 +
c_1^2 \left[ k_x - \frac{d}{4e} E(t) \right]^2 + M^2(t) \; .
\label{omega}
\ee

To achieve the correct commutation relation between $X$ and $\partial_t X$,
the mode functions $f$ should be normalized by the condition
\be
f_{\k} \dot{f}^*_{\k} - \dot{f}_{\k} f^*_{\k} = \frac{i}{V} \; ,
\label{W}
\ee
where $V$ is the total two-dimensional volume. 
At some initial time $t=t_i$, before the pulse, we have $E(t_i) = 0$ and
$M(t_i) = M_0$, so we can use the usual plane-wave exponentials 
as initial conditions for $f_{\k}(t)$. Equivalently,
\ba
f_{\k}(t_i) & = & [2\omega_{\k}(t_i) V]^{-1/2} \; , \label{ini1} \\
\dot{f}_{\k}(t_i) & = & - i \omega_k f_{\k}(t_i) \; , \label{ini2}
\ea
where $\omega_{\k}^2(t_i) = c_1^2 k^2 + M_0^2$.

As discussed in the preceding sections, to include the effect of the
vortex-antivortex potential, the frequency $M(t)$ in these formulas is 
taken to be the average frequency along the vortex path. We note also
that eq. (\ref{f}) with the initial conditions (\ref{ini1})--(\ref{ini2})
can be used for numerical studies of vortex transport \cite{num}.

Substituting (\ref{four}) in (\ref{Jx}), and assuming that the quantum
state of the system is the vacuum of the operators $\alpha$ and 
$\beta$, we obtain the average current as
\be
\langle J_x(t) \rangle = 2 c_1^2 \sum_{\k} 
\left[ k_x - \frac{d}{4e} E(t) \right] |f_{\k}|^2 \; .
\label{Jave}
\ee
The integral of this over time gives the average total vortex number 
transported around the cylinder (i.e. in the $x$ direction), per unit 
length in the $y$ direction. So, it measures the efficiency of the
amplitude transfer between the basis states of the qubit.

The second quantity of interest is a measure of the residual excitation
left in the system after the pulse. It is simply the total
average number of vortices left at some final time $t=t_f$ and is 
obtained
as the sum $\sum_{\k} n_{\k}(t_f)$ of the occupation numbers of all the 
individual modes. These occupation numbers, as functions of time, are
given by
\be
n_{\k}(t) = \frac{V}{2\omega_{\k}(t)} \left[
|\dot{f}_{\k}(t)|^2 + \omega^2_{\k}(t)|f_{\k}(t)|^2 \right] - \half \; .
\label{nk}
\ee
Our next goal will be to show that a sufficiently slow, adiabatic change
in $E$ and $M$ can lead to a sizable transport of the vortex number, while
leaving $n_{\k}(t\to\infty)$ exponentially suppressed.

\section{Adiabatic limit}
When the frequency (\ref{omega}) for each mode changes with time 
slowly (adiabatically), i.e. 
\be
|\partial_t \omega_{\k}| \ll \omega_{\k}^2 \; ,
\label{adiab}
\ee
the adiabatic theorem \cite{adiab}
guarantees that particle production will be absent in
any finite order in $|\partial_t \omega_{\k}| / \omega_{\k}^2$. In other
words, $n_{\k}(t)$, which was zero initially, will be zero at $t\to\infty$
to exponential accuracy. 
The mode functions can then be approximated by WKB-type expressions:
\be
f_{\k}(t) \approx [2\omega_{\k}(t) V]^{-1/2}
\exp\left\{ -i \int_{t_i}^t \omega_{\k}(t') dt' \right\} \; .
\label{WKB}
\ee
Substituting this into (\ref{Jave}), we obtain the average current as
\be
\langle J_x(t) \rangle \approx \frac{c_1^2}{V} \sum_{\k} 
\frac{1}{\omega_{\k}(t)} \left[ k_x -  \tE(t) \right] \; ,
\label{J1}
\ee
where we have introduced notation
\be
\tE(t) \equiv \frac{d}{4e} E(t) \; .
\label{tE}
\ee

The summand in (\ref{J1}) depends on $k_x$ only in combination 
$k_x -  \tE$, and it is odd in that combination. So, if we could replace the
sum over $k_x$ by an integral and make a shift of the integration variable,
we would prove that (\ref{J1}) is zero. There are, however, two obstructions
to this procedure. First, the integral needs an ultraviolet
regularization, which we take to be symmetric in $k_x$, not in $k_x -  \tE$.
As a result, the far ultraviolet modes contribute a finite amount 
proportional to
$\tE$. Second, the difference between the sum and the integral results in
a correction, which is periodic in $\tE$ with period $2\pi/ L_x$, where
$L_x$ is the circumference of the cylinder. This correction vanishes
when $\tE$ is an integer multiple of $\pi / L_x$, but is finite otherwise.
It will be important for us to 
understand the structure of this correction.

To make the argument more transparent, let us consider 
the case of a weak electric field,
$|\tE| \ll 2\pi / L_x$, so that we can expand the expected periodic 
correction
in $\tE$. To simplify things even further, we will also assume that
\be
b' = \frac{M L_x}{2c_1} \gg 1 \; .
\label{b'}
\ee
Then, we can expand $1/\omega_{\k}$ in (\ref{J1}) in $\tE$, to obtain
\ba
\lefteqn{
\frac{1}{\omega_{\k}} ( k_x -  \tE ) }
\nonumber \\
& & = (k^2 + M^2)^{-1/2} \left\{ k_x
- \frac{c_1^2 k_y^2 + M^2}{c_1^2 k^2 + M^2} \tE + O(\tE^2) \right\} \; .
\label{expand}
\ea
The first term in the braces gives zero upon summation over $k_x$ (assuming
a symmetric ultraviolet cutoff), so we have
\be
\langle J_x(t) \rangle \approx -\frac{c_1^2 \tE }{V} 
\sum_{k_y} m^2 \sum_{k_x} \frac{1}{(c_1^2 k_x^2 + m^2)^{3/2}} 
+ O(\tE^2) \; , 
\label{J2}
\ee
where
\be
m^2 = c_1^2 k_y^2 + M^2 \; .
\label{m2}
\ee

To compute the sum over $k_x$ in (\ref{J2}), we use the representation
\ba
\lefteqn{\sum_{n=-\infty}^{\infty} \frac{1}{(\pi^2 n^2 + b^2)^{3/2}} }
\nonumber \\ 
& & = \frac{2}{\pi b^2} \left( 1 + \int_b^{\infty} 
\frac{\omega d\omega} {\sqrt{\omega^2 - b^2}~\sinh^2\omega}
\right) \; ;
\label{rep}
\ea
in our case
\be
b = \frac{m L_x}{2c_1} \; .
\label{bdef}
\ee
Under the condition (\ref{b'}), $b$ is large, and we can expand the integral
in (\ref{rep}) in $e^{-2b}$. We obtain
\ba
\lefteqn{ \sum_{k_x} \frac{1}{(c_1^2 k_x^2 + m^2)^{3/2}} }
\nonumber \\
& & = \frac{L_x}{\pi c_1 m^2} \left[ 1 + 2\sqrt{\pi b} e^{-2b} 
+ O(e^{-4b}) \right]
\label{sum_kx}
\; .
\ea

If we were to neglect the discreteness of modes, i.e.
replace the sum on the left-hand side of (\ref{sum_kx}) with an integral,
we would obtain only the first term in the bracket. According to
(\ref{J2}), the average current due to this term is proportional
to $E$, with no other time-dependent factors 
(the factors of $m^2$ cancel out). 
In the absence of sources of dissipation, such as 
pair-produced quasiparticles or real (non-virtual) vortices, we have
$E = -(1/c) \partial_t A$, so the integral of $E$ over time is zero.
Thus, the total vortex number transported around the cylinder is determined
entirely by the exponential correction in (\ref{sum_kx}).

The remaining sum over $k_y$ can be replaced by an integral and easily
evaluated, noting that, at large values of $b'$, $e^{-2b}$ 
confines $k_y$ to rather small values: $c_1^2 k_y^2 \sim M^2 / b'$.
Thus, in the limit (\ref{b'}) we finally obtain
\be
\int dt \langle J_x(t) \rangle \approx
-\frac{1}{\pi} \int dt \tE M \exp(-M L_x / c_1) \; .
\label{int}
\ee
This quantity has dimension of inverse length, as it gives the transported
vortex number per unit length of the cylinder.

The exponential factor in (\ref{int}) coincides with the one obtained in
sect. \ref{sect:env} by means of instanton calculus. Moreover, the
periodicity with respect to $\tE$, noted after eq. (\ref{tE}), has 
a simple explanation in the instanton picture. Indeed, from (\ref{eqm2})
we see that the product $\tE L_x$ equals (up to a constant) the ``solenoidal flux''
$\int a_x dx$. So, for $\tE \neq 0$ the amplitudes due to a single
instanton and a single antiinstanton acquire opposite phases, equal 
in magnitude to $|\tE L_x|$. In
the limit (\ref{b'}), when the instanton gas is dilute, the average vortex
current is then proportional to
\be
\exp(-M L_x / c_1) \sin \tE L_x \; ,
\ee
which has precisely the same periodicity as that deduced from eq. 
(\ref{J1}).
We stress that in the dual calculation these results appear as 
a consequence of the quantization of vortex modes in a confined geometry, 
without any reference to instantons. So, the present calculation is 
complementary to the instanton calculation of sect. \ref{sect:env}.

Note that eq. (\ref{int}) determines also the value of the preexponent. As
we have already discussed, unlike the exponent, the preexponent may
in certain cases be affected by fermions at the vortex core.
Any such modification will be absent from eq. (\ref{int}).
We will, however, continue to use this equation, since our main conclusions
are based on the value of the exponential factor.

The exponential suppression seen in (\ref{int})
is of entirely different origin---and, hence, generally of different 
magnitude---than the adiabatic suppression of $n_{\k}$. To remove 
the exponential suppression in (\ref{int}) altogether, we need, 
at some time $t$ during the pulse, to go just outside the limit
(\ref{b'}), i.e. achieve the condition
\be
M(t) = c_1 / L_x \; .
\label{Mt}
\ee
(At this point, the $O(e^{-4b})$ terms
in (\ref{sum_kx}) will become important.) Eq. (\ref{Mt}) will not 
jeopardize adiabaticity provided that the characteristic timescale $t_p$
of the pulse (e.g. the ramp time) satisfies
\be
t_p / 2\pi \gg 1 / M(t) = L_x / c_1 \; .
\label{tp}
\ee
On the other hand, one
can choose parameters of the device so that, before and after the pulse, 
the probability of random ``errors'', i.e., spontaneous transitions between
the basis states, is vanishingly small. (The amplitude of such spontaneous
transitions is proportional to  $\exp(-M_0 L_x / c_1)$, where $M_0$ is the
unperturbed initial mass.)

\section{Numerical estimates} \label{sect:num}
Let us summarize various conditions we have obtained so far on the parameters
of the system. One is the inequality (\ref{tp}), which is 
the condition that a significant (order 1) vortex number transport is
compatible with the adiabatic suppression of production of real vortex 
pairs. Another is the condition of exponential 
suppression of spontaneous tunneling transitions (``errors''):
\be
L_x \gg \lambda \equiv c_1 / M_0    \; .
\label{lam}
\ee
The quantity $\lambda$ here, of the dimension of length, is analogous
to the Compton wavelength of an elementary particle.

To (\ref{tp}) and (\ref{lam}), we should add the condition that pair 
production of quasiparticles is also exponentially suppressed 
(since quasiparticles, like vortices, lead to decoherence\footnote{
A quasiparticle pair can detect passage of flux through Aharonov-Bohm
scattering. Although it remains to calculate how strong this effect
actually is, we nevertheless impose the condition of negligible 
pair production below.}) and the condition
that the coherence length of the superconductor remains smaller than 
$L_x$, so that vortices remain well defined.
These two conditions should be written so as to take into account the
reduction of the superconducting gap $\bar{\Delta}$, 
and the corresponding increase
in the coherence length $\bar{\xi}$, during the pulse:
\ba
t_p/ 2\pi & \gg & \hbar/\bar{\Delta}(t) = \frac{\hbar\bar{\xi}(t)}{\xi \Delta}
\; , \label{Dbar} \\
L_x & \gg & \bar{\xi}(t) \; . \label{xibar}
\ea
Quantities without the bar denote the unperturbed values, i.e. those
before and after the pulse. We now obtain estimates for $M_0$, $c_1$, and 
$\lambda$.

To obtain these estimates we use the Ginzburg-Landau (GL) 
energy of superconductor
\be
H_{\rm GL} = d \int d^2 x \left[
\zeta |(\nabla + ig\A)\psi|^{2} - |r| |\psi|^2 + \frac{s}{2} |\psi|^4 \right] 
\; ,
\label{HGL}
\ee 
where $\zeta$, $r$, and $s$ are parameters. 
Although this expression is strictly applicable only near critical 
temperature, we will use it at temperatures close to absolute zero,
where our device is supposed to operate, since all we need are 
order-of-magnitude estimates.

From the GL energy (\ref{HGL}), $M_0$, which was defined as half of the
threshold frequency required to produce a real vortex-antivortex pair,
can be estimated as
\be
\hbar M_0 \sim 2\pi d \zeta |\psi_0|^2 \; .
\label{M0}
\ee
Here $\psi_0$ is the order parameter in the absence of current. This
is an estimate for the ``core'' frequency. The long-range interaction between
a vortex and an antivortex will contribute an additional amount, enhanced
relative to (\ref{M0}) by $\ln L_x /\bar{\xi}$. However, since for
the values of the parameters that we use below this logarithm is of order
one, the estimate (\ref{M0}) will be sufficient for our purposes.
It can be rewritten using the London screening length $\delta$:
\be
\hbar M_0 \sim \frac{e^2 d }{16 \alpha_{\rm EM}^2 \delta^2} \; ,
\label{M0_est}
\ee
where $\delta^{-2} = 32\pi e^2 \zeta |\psi_0|^2 / \hbar^2 c^2$, and
$\alpha_{\rm EM}$ is the fine structure constant. (In thin films,
$\delta$ determines the strength of the London current but not the actual
screening length of the vortex magnetic field.)

We now turn to estimating the limiting speed 
$c_1$, or equivalently the vortex inertial mass (\ref{m}). 
There are two types of contributions to $m$ \cite{micro,Sonin&al}: 
from the electric field caused by the vortex motion and from core fermions.
We have discussed the mass due to core fermions in  
sect. \ref{sect:env} and have seen that for vortex tunneling,
characterized by short timescales, the effect is not as large as for slow
motions: on a short timescale the vortex cannot transfer a large
longitudinal momentum to the fermion subsystem.
In particular, the core contribution to the mass does not affect the value 
of the instanton exponent.
Accordingly, for our estimates we take $m$ to be given by the small 
electromagnetic mass of the vortex
\be
m \sim \frac{\hbar^2 d}{16 e^2 \xi^2} \; .
\label{inert}
\ee
(See refs. \cite{Suhl,Sonin&al,thermal} for how this estimate can be obtained.)
This estimate results in a large value of $c_1$:
\be
c_1 \sim (\xi / \delta) c \; .
\label{c1_est}
\ee
The vortex's ``Compton wavelength'' can be now obtained from (\ref{lam}):
\be
\lambda \sim  \frac{16\alpha_{\rm EM} \xi\delta }{d} \; .
\label{lam_est}
\ee

For numerical estimates, we use the following values: $d = 10$ nm,
$\xi = 30$ nm, and $\delta = 100$ nm. Then, according to (\ref{lam_est}),
$\lambda \sim 35$ nm, and (\ref{lam}) suggests that to suppress 
spontaneous transitions it is sufficient to use $L_x$ of order 1 $\mu$m.
Comparing (\ref{Mt}) and (\ref{lam}), we see that in this case
the pulse will need
to reduce the vortex frequency $M$ by a factor
\be
\frac{M_0}{M(t)} = \frac{L_x}{\lambda} \sim 30 \; .
\label{fac}
\ee
Because $M(t)$ depends quadratically on the gap $\bar{\Delta}(t)$ 
(cf. (\ref{M0})), eq. (\ref{fac}) 
corresponds to a reduction in $\bar{\Delta}$, and
an increase in $\bar{\xi}$, by a factor of order 5. 
For a superconductor with a
quasiparticle gap $\Delta= 10$ K, this turns the condition 
(\ref{Dbar}) into
\be
t_p/ 2\pi \gg 4\times 10^{-12} ~{\rm s} \; .
\label{tp1}
\ee
Meanwhile, the competing condition (\ref{tp}) is
\be
t_p/ 2\pi \gg 10^{-14} ~{\rm s} \; .
\label{tp2}
\ee
So, for the above values of the parameters, adiabaticity with respect 
to quasiparticle production is a stronger condition than adiabaticity 
with respect to production of vortices. Finally, the condition 
(\ref{xibar}) can be easily verified.

\section{Conclusion}
Our main result is the demonstration that for topological transitions,
involving transport of magnetic flux via vortices,
there exists a potentially useful adiabatic limit. 
In that limit, the probability
to produce real vortex or quasiparticle pairs is exponentially small
(as might be expected), but the transport of the vortex number due to
{\em virtual} vortices can in principle be of order one.

Our choice of the cylindrical geometry was motivated by the ease of
imposing boundary conditions on the vortex field in the dual formulation.
We expect the result to apply to other geometries, which may also be 
easier to fabricate.

We have presented two calculations of the leading exponential factor
in the tunneling amplitude: one based on instantons, and the other
based on the dual formulation, in which vortices are described by
a fundamental quantum field. The results have been shown to agree.

In addition, the instanton approach has allowed us to estimate the
effects of the core fermions on vortex tunneling. The process at hand,
i.e. an adiabatic transition in which no vortices are
produced in the final state, turns out to have a very short
(Euclidean) timescale, resulting in only a small transfer of 
longitudinal momentum to the fermions. As a consequence, the effective
inertial mass of the vortex during tunneling
is much smaller than the known estimates corresponding to
slow motions. We have also discussed the effect of the vortex-antivortex
potential and have found that, for calculating the leading exponential
factor, the position-dependent vortex mass can be replaced with its average
over the vortex path.

Our calculations were done under the condition that the wire width $L_x$
(the circumference of the cylinder) exceeds the GL coherence length $\xi$,
so that vortices remain well defined. (This is the limit opposite to
that of an ultra-thin wire, where tunneling is due to quantum phase
slips.) Our results suggest that for $L_x$ in the micron range, induced
vortex tunneling may be experimentally observable.

Moreover, since the adiabatic limit, i.e. a slow switching on and off of 
the current, eliminate the most obvious sources of decoherence, namely,
pair production of vortices or quasiparticles, the process may
be suitable for forming quantum superpositions of flux states.
There are other possible sources of 
decoherence, in particular those due to components of the environment,
such as nuclear or impurity spins, that are sensitive to magnetic fields 
caused by persistent currents, but we do not expect the situation here to be 
worse than for the conventional SQUID designs \cite{Orlando&al}.

The author thanks Dane Bass and Albert Chang for discussions, and M. Stone
and G. Volovik for correspondence on the physics of core fermions.

\end{document}